\documentstyle[epsf]{mn}
\def\ion#1#2{{\rm #1}%
\ifmmode{\mathchoice{\scriptstyle}{\scriptstyle}
{\scriptscriptstyle}{\scriptscriptstyle}{\rm\uppercase{#2}}}%
\else$\,\scriptstyle\rm\uppercase{#2}$\fi}
\def\HII{{\ion{H}{II}}}
\def\refeq#1{{(\ref{#1})}}
\def\id{{\rm\,d}}                                
\def\etal{{et~al.}}
\def\eg{{e.g.}}
\def\ie{{i.e.}}

\def\cf{{cf.}}
\def\unit#1{\,{\rm {#1}}}
\def\umult#1#2{\ifx#2\unit\def\tmpa{\umulti#1#2}\else\toks0=\expandafter{#2}%
\edef\tmpa{\noexpand\umulti\noexpand#1\the\toks0}\fi\tmpa}
\def\umulti#1#2#3{\ifx#2\unit\unit{#1#3}\else%
\message{\noexpand\umult error: must precede unit or \noexpand\unit}%
\unit{#1}#2#3\fi}

\def\kms{\unit{km\,s^{-1}}}
\def\secnd{\unit{s}}
\def\yr{\unit{yr}}
\def\gram{\unit{g}}
\def\cm{\unit{cm}}

\def\parsec{\unit{pc}}

\def\Msun{\unit{M_\odot}}

\def\Kelv{\unit{K}}
\def\subrm#1{_{\rm #1}}
\catcode`|=\active \let|=\subrm
\def\ee{\protect\pee}
\def\pee#1{\ifmmode{\times10^{#1}}\else$\times10^{#1}$\fi}
\newcommand{\UCHIIR}{{UCH{\sc ii}R}}
\newcommand{\domust}{{\dot{\mu}_\star}}
\newdimen \figwidth
\newif\ifref
\reffalse
\ifref
\figwidth \textwidth
\else
\fi
\title[Clumpy ultracompact H{\small\it\,II} regions I]
{Clumpy ultracompact H{\Large\bf\,II} regions I:\\ Fully supersonic
wind-blown models}
\author[J.E. Dyson, R.J.R. Williams and M.P. Redman]
{J.E. Dyson, R.J.R. Williams and M.P. Redman\\
Department~of~Physics~and~Astronomy,
University~of~Manchester, Oxford~Road, Manchester M13~9PL}

\date{Received **INSERT**; in original form **INSERT**}
\pagerange{\pageref{firstpage}--\pageref{lastpage}}

\begin{document}
\label{firstpage}
\maketitle

\begin{abstract}
We propose that a significant fraction of the ultracompact \HII\
regions found in massive star-forming clouds are the result of the
interaction of the wind and ionizing radiation from a young massive
star with the clumpy molecular cloud gas in its neighbourhood.
Distributed mass loading in the flow allows the compact nebulae to be
long-lived.  In this paper, we discuss a particularly simple case, in
which the flow in the \HII\ region is everywhere supersonic.  The line
profiles predicted for this model are highly characteristic, for the
case of uniform mass loading.  We discuss briefly other observational
diagnostics of these models.
\end{abstract}

\begin{keywords}
hydrodynamics -- stars: mass-loss -- ISM: structure --
\HII\ regions -- radio lines: ISM
\end{keywords}

\section{Introduction}

The formation of massive stars is an important, but vexed, question,
largely because such stars disrupt their natal environment by their
powerful winds and ultraviolet radiation fields.  The ultracompact
\HII\ regions (\UCHIIR), which are found deep within molecular clouds,
may provide significant information on the properties and environment
of the very youngest massive stars.  \UCHIIR\ are highly obscured at
optical wavelengths and most information on them derives from infrared
and high frequency radio data.  A comprehensive overview of their
properties is given by Churchwell (1990).  They have high emission
measures ($\langle n|e^2 \rangle L \ga 10^7\cm^{-6}\parsec$) and small
scale sizes ($L\la 10^{17}\cm$) and thus high r.m.s.\ electron
densities ($\langle n|e^2\rangle^{1/2} \ga 10^4\cm^{-3}$).  The
ionizing photon supply rates necessary to maintain the observed
ionization are in the approximate range
$10^{44}$--$10^{49}\secnd^{-1}$, corresponding roughly to ZAMS
spectral types B2--O5 for single star ionization.  There is, however,
considerable evidence from comparison of IR and radio data, that
groups of stars may be embedded within a given \UCHIIR\ and also that
the dust content of the region may absorb nearly all the UV photons
generated by the stars (Kurtz, Churchwell \& Wood 1994).
Consequently, the assignment of a unique spectral type to the exciting
source is an extremely uncertain procedure.

\UCHIIR\ have a wide range of observed morphologies.  Churchwell (1990)
defines 5 main morphological types: cometary ($\sim 20$ per cent),
core-halo ($\sim 16$ per cent), shell ($\sim 4$ per cent), irregular
or multiply peaked ($\sim 17$ per cent) and spherical and unresolved
($\sim 43$ per cent).  Most theoretical attention has been given to
cometary regions (\eg\ Van Buren \& Mac Low 1992).  They can be
modelled fairly convincingly as the steady-state partially ionized
structures behind bow shocks driven into molecular cloud material by
the winds of moving stars.

Although Mac Low \etal\ (1991) suggest that morphologies other than
cometary (in particular core-halo) can be explained as cometary
structures viewed along the axis of symmetry, it seems likely that
other models need consideration.  For example, Kurtz, Churchwell \&
Wood (1994) give evidence for a size-density relationship for
spherical or unresolved regions.  This is neither present nor would be
expected for cometary \UCHIIR.  Dyson (1994) suggested that a natural
explanation for at least some \UCHIIR\ arises from the interaction
between an early type star and a very clumpy, as opposed to relatively
homogeneous, molecular cloud.  Molecular clouds are well established
to be clumpy on length scales down to the limits of observational
resolution.  Clumps can act as localised reservoirs of gas which can
be injected into the surroundings by photoionization and/or
hydrodynamic ablation.  The continuous injection of material into a
\HII\ region can lead to a quasi steady-state configuration which is
bounded by a recombination front (RF) as opposed to the ionization
front (IF) which bounds standard \HII\ regions.  This particular
configuration avoids the expansion problems associated with a
`classical' \HII\ region model for \UCHIIR, which would expand on a
timescale far shorter than the characteristic $10^{5-6}\yr$ lifetime
of \UCHIIR\ estimated from the statistics of \UCHIIR\ association with
early type stars (Churchwell 1990). \HII\ regions bounded by
recombination fronts can be held in equilibrium by gas pressure or
recoil pressure depending on whether the flow exits sub- or
super-sonically.

We discuss here a particularly simple model where a star with both a
UV radiation field and a powerful hypersonic wind interacts with
clumpy material and show that it may explain the structure of
shell-like \UCHIIR.  Further papers in this series will investigate in
detail the full range of models outlined by Dyson (1994).  A future
key test of all these models will be the observational determination
of the kinematics of ionized gas, molecular gas and neutral hydrogen
associated with such \UCHIIR.

Non-steady models with photoionization-induced mass injection have
also been suggested by Lizano \etal\ (1995).  One major difference
between their models and those given here and by Dyson (1994) is that
Lizano \etal\ assume a very specific model for the mass injection
process which leads to a close coupling between the spatial and
temporal behaviour of the mass injection and the radiation field.  We
treat mass injection as a free parameter and assume here that it is
constant in space and time.  Dyson (1994) has discussed some of the
factors involved in the mass injection process and emphasised that
both hydrodynamics and photoinjection may play roles.  It is clear
that both approaches need investigation.

\section{\UCHIIR\ as wind driven flows in clumpy molecular clouds}

In this initial paper, we assume that the hypersonic wind from a
single star blows into a clumpy molecular cloud and that mass
injection into the flow occurs at a constant volume rate $\dot{q}$
near to the star.  Mass injected into a fast wind will slow the wind
down.  The frictional energy dissipated in this process may be
radiated away by, for example, enhanced radiative losses in boundary
layers (Hartquist \& Dyson 1993).  We assume that this occurs here and
that the flow is photoionized by the stellar radiation field and
remains isothermal at the usual photoionization equilibrium
temperature $T\simeq 10^4\Kelv$.  A more detailed discussion of the
isothermality of mass loaded winds is given elsewhere (Williams,
Hartquist \& Dyson 1995).

We assume that mass loading occurs only in the ionized region and that
the flow always remains supersonic.  We neglect the possibility of
global shocks in the flow produced by mass injection.  This will be a
valid assumption provided that the Mach number in the ionized region
is not predicted to fall below about 2 (Williams \etal\ 1995).  The
recombination front which then occurs has supersonic inflow and
outflow (\cf\ Newman \& Axford 1968).  A discussion of the
consequences of global shocks and subsequent transonic flow for RF
structures is given elsewhere (Williams \& Dyson, in preparation).

For steady supersonic flow with uniform mass loading, the mass and
momentum conservation equations are respectively
\begin{equation}
\rho u = {1\over3}\dot{q}r\label{e:masscons}
\end{equation}
and
\begin{equation}
4\pi r^2\rho u^2 = \dot{M}_\star v_\star \equiv \domust,\label{e:momm}
\end{equation}
where $u$ and $\rho$ are respectively the flow velocity and density,
$r$ is the radial coordinate and $\dot{M}_\star$ and $v_\star$ the
stellar wind mass loss rate and velocity.  We have assumed that the
mass flux in the flow is dominated by gas ablated from the clumps.
Then from equations~\refeq{e:masscons} and~\refeq{e:momm},
\begin{equation}
u = {3\domust\over4\pi\dot{q}r^3};\quad
\rho = {4\pi\dot{q}^2r^4\over9\domust}.\label{e:struct}
\end{equation}
Clearly, radiation such as bremsstrahlung and recombination lines,
which have emissivities proportional to $\rho^2$, will be dominated by
the outer regions of the flow.

The RF occurs at a radius, $R|R$, where the photon output rate
balances the recombination rate.  We assume an effective stellar
output of $S_\star\,\mbox{photons}\secnd^{-1}$ in the Lyman continuum
(which may be only a small fraction of that actually generated by the
star because of absorption by dust).  We assume (and justify later)
that if mass injection is a result of photoionization, it needs only a
negligibly small fraction of the available UV.  If the observed nebula
is ionization bounded, then it must be powered by a Lyman continuum
flux of
\begin{equation}
S_\star =\int_0^{R|R}4\pi {r'}^2 n^2 \beta_2 \id{r'},\label{e:front}
\end{equation}
where $n$ ($\equiv \rho/\bar m$) follows from
equation~\refeq{e:struct}, $\bar m$ is the mean mass per nucleon and
\hbox{$\beta_2$ ($=2\ee{-13}\cm^3\secnd^{-1}$)} is the hydrogen recombination
coefficient.  Thus from equation~\refeq{e:front}
\begin{equation}
\dot{q} = \left(891\over64\pi^3\right)^{1/4}
	\domust^{1/2} \bar{m}^{1/2}S_\star^{1/4}\beta_2^{-1/4}R|R^{-11/4}.
	\label{e:massrate}
\end{equation}
We treat $\dot{q}$ as a free parameter since it is presumably
determined by factors such as the density of mass loading centres and
the mode of ablation (Section 1).

Equation~\refeq{e:front} implies that we can neglect any appreciable
contributions to the radio emission from the mass loading centres
themselves (\eg\ from dense ionized clump surfaces), and so the
measured r.m.s.\ electron density is that of the flow and is given by
\begin{equation}
\langle n|e^2\rangle^{1/2} =
\left[{3\over R|R^3}\int_0^{R|R}
\left(16\pi^2\dot{q}^4\over81\domust^2\bar{m}^2\right)
r^{10}\id{r}\right]^{1/2}.
\end{equation}
{}From equation~\refeq{e:massrate} therefore
\begin{equation}
\langle n|e^2\rangle^{1/2} =
\left(3\over 4\pi\right)^{1/2} S_\star^{1/2} R|R^{-3/2}\beta_2^{-1/2}.
\label{e:modelne}
\end{equation}
The Mach number just before the RF is
\begin{equation}
M|R \equiv {u|R\over c|i} = {3\domust\over4\pi\dot{q}R|R^3c|i}
\label{e:modelmach}
\end{equation}
Inserting characteristic values
$S_{48} \equiv S_\star/10^{48}\secnd^{-1}$;
$\dot\mu_{28} \equiv \domust/10^{28}\gram\cm\secnd^{-2}$;
$R_{17} \equiv R|R/10^{17}\cm$;
$c|i =10\kms$ and $\bar{m} = 2\ee{-24}\gram$,
equations~\refeq{e:modelne} and~\refeq{e:modelmach} become
\begin{eqnarray}
\langle n|e^2\rangle^{1/2} &= &
3.5\ee{4} S_{48}^{1/2} R_{17}^{-3/2}\cm^{-3}\label{e:densmod} \\
M|R & = &
0.78 S_{48}^{-1/4}R_{17}^{-1/4}\dot\mu_{28}^{1/2}.\label{e:machmod}
\end{eqnarray}
The mass injection rate required is
\begin{equation}
\dot{q} =
2.4\ee{-30}\dot\mu_{28}R_{17}^{-3}M|R^{-1} \gram\cm^{-3}\secnd^{-1}.
\end{equation}

The Mach number at the recombination front from
equation~\refeq{e:machmod} is smaller than required, for the values of
$S_{48}$ and $R_{17}$ derived from observation, unless the momentum
input (for the observed $S_{48}$) is considerably larger than that
found for field OB stars.  Dust significantly reduces the fraction of
$S_\star$ which ionizes the nebula (and thus increases the effective
$\mu_{28}^2/S_{48}$); however, in the sample of Kurtz \etal, the
fraction of the stellar luminosity absorbed by dust is less for the
earlier type stars which would have stronger winds.  The presence of
multiple stars within a single \UCHIIR\ may also have an effect.

It is certain that the flow will have significant density
inhomogeneities.  These can decrease the radius of the recombination
front for given mass fluxes and ionizing luminosity.  We assume that
the inhomogeneities have a large density contrast, and that the dense
gas dominates both the emission and the mass flux compared to the more
tenuous surrounding gas.  In this case, the Mach number of the
recombination front [equation~\refeq{e:machmod}] scales as $M|R\propto
f^{-1/4}$, where $f$ is the volume filling factor of the dense
regions; the mean density of the flow at any radius is $f^{1/2}$ times
smaller than that derived (for observed $S_\star$ and $R|R$) by
assuming a smooth flow.  High resolution images of \UCHIIR\ do indeed
show a strongly clumped structure (\eg\ Kurtz \etal\ 1994).  In the
present paper we assume that the dense gas has a large ($\simeq 1$)
covering factor as seen from the central star, and that the dense gas
is well coupled to the global flow.  Exploration of these assumptions
demands a treatment of intermediate scale structure in the flow
(Dyson, Hartquist \& Williams, in preparation).  We have also assumed
here that the filling factor is independent of radius.

To see whether the required mass injection is plausible and also
compatible with the assumptions in the model we treat the very special
case where the mass injection comes from the surfaces of pressure
confined self-gravitating isothermal clumps (Dyson 1968; Kahn 1969).
Kahn (1969) showed that the mass loss rate from such a photoionized
globule had the remarkable property of being dependent only on the
sound speed $c|n$ in the neutral gas of the globule.  We adopt his
mass loss rate from a globule $\dot{m}|g \simeq 10^{20}
c_3^4\gram\secnd^{-1}$, where the sound speed in the cloud is $c|n =
0.3c_3\kms$ to allow for its being cold and molecular.  The number of
clumps $N|c$ required within the \UCHIIR\ is thus $N|c \simeq
(4\pi/3)R|R^3\dot{q}/\dot{m}|g \simeq 130
R_{17}^{1/4}f^{1/4}S_{48}^{1/4}\dot\mu_{28}^{1/2}c_3^{-4}$ which
appears plausible.  The fraction, $\phi$, of photons used up in the
photoinjection process itself is $\phi = (4\pi
R|R^3\dot{q}/3\bar{m}S_\star) \simeq 6\ee{-3}
R_{17}^{1/4}f^{1/4}S_{48}^{-3/4}\dot\mu_{28}^{1/2}$, which is
consistent with the assumptions made.

Mass loss driven by ionization from gravitationally confined clumps is
just one of a wide range of possibilities.  Hydrodynamical ablation
may play an important role in releasing gas from the clumps, for
instance, as may magnetic fields in maintaining inhomogeneities within
a global flow.  There must be several tens of mass loading clumps
within the \UCHIIR, if our continuum approximation is to be reasonably
valid.  There is strong observational evidence for such clumping in
the emitting gas, ten or more individual emission peaks being visible
in the better-resolved images in the survey by Kurtz, Churchwell \&
Wood (1994).

This characteristic number of clumps implies a space density of $\sim
10^6 \parsec^{-3}$.  The clumps may be related to the partially
ionized globules (PIGS) and protoplanetary disks around low mass stars
(proplyds, O'Dell \& Wen 1994) observed distributed in Orion nebula.
The peak number density of stars in the low mass stellar cluster
around the Trapezium OB stars is at least $5\ee4 \parsec^{-3}$
(McCaughrean \& Stauffer 1995).  While the observed velocity structure
of these sources (Massey \& Meaburn 1995) may well be similar to that
of the mass-loading clumps in \UCHIIR, the Orion nebula is a far older
\HII\ region than those discussed here.  It is likely that in the near
vicinity of a very young massive star, there will also be many, more
transient, local density enhancements which have not quite passed the
threshold for star formation.  In $10^5\yr$, between 1 and $10\Msun$
of gas will pass through a near-sonic recombination front, suggesting
a mean density of clump gas in the region ($3\ee5-3\ee6\cm^{-3}$),
comparable to the mean mass density of the proplyd stars in Orion.

It is also interesting to compare the Keplerian velocity at the edge
of an \UCHIIR\ (around $1\kms$) and the relative velocity necessary to
transit the \UCHIIR\ in its lifetime (around $0.3\kms$) to the mean
velocity dispersion of the Trapezium stars (around $3\kms$).  A
significant flux of new clumps will enter the \UCHIIR\ during its
lifetime, either through random motions or systematic infall; the
density of clumps within the region may also be significantly enhanced
by gravitational focussing effects.

\section{A simple model \UCHIIR}

In order to satisfy the requirement that no global shocks occur in the
\UCHIIR, we require $M|R\ga 2$ (Section 2), \ie\
$(fS_{48}R_{17}/\dot\mu_{28}^2)\la 0.03$.  We therefore choose as an
illustrative set of parameters $f=0.02$, $S_{48} = 1$, $R_{17} = 1$,
$\dot\mu_{28} = 1$.   We then find the observed r.m.s.\ density,
$\langle n|e^2\rangle^{1/2} \simeq 3.5\ee4\cm^{-3}$ (equivalent to an
r.m.s.\ density in the smoothed flow of $5\ee3\cm^{-3}$),
$\dot{q}\simeq 1.2\ee{-30}\gram\cm^{-3}\secnd^{-1}$, $\phi=2.3\ee{-3}$
and $M|R=2.2$.

\begin{figure}
\centering
\mbox{\epsfxsize=\figwidth\epsfbox{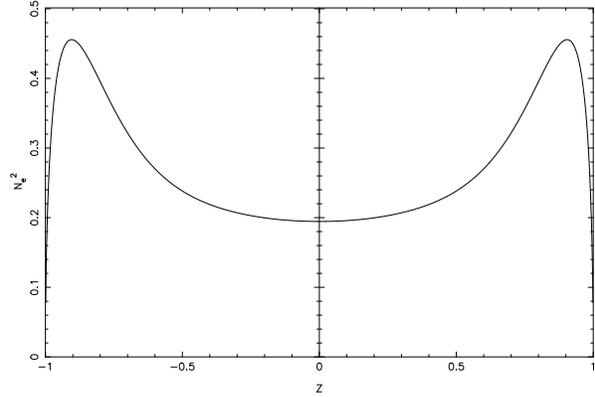}}
\caption{Emission measure as a function of $z$ (fractional offset from
the centre of the \UCHIIR) for $M|R=2.2$, normalized to unit total
emission.}
\label{f:emplot}
\end{figure}
\begin{figure*}
\centering
\begin{tabular}{cc}
\multicolumn{1}{l}{(a)} &
\multicolumn{1}{l}{(b)} \\
\epsfxsize=\textwidth\divide\epsfxsize by 2
\mbox{\epsfbox{lprof01.ps}} &
\epsfxsize=\textwidth\divide\epsfxsize by 2
\mbox{\epsfbox{lprof07.ps}}
\end{tabular}
\caption{Line profiles at (a) $z=0.1$ and (b) $z = 0.7$, normalized
to unit total emission across the nebula (when integrated over the
area of the nebula in units of $R|R$ and projected velocity in units
of $c|i$), for $M|R=2.2$.}
\label{f:lineprof}
\end{figure*}
In Figure~\ref{f:emplot}, we show the emission measure, ${\rm EM} =
\int n^2 \id{l}$, as a function of offset $z$ from the star (where $z$
is given in units of $R|R$).  The total emission is normalized so that
\begin{equation}
\int_0^1 {\rm EM}(z) 2\pi z \id{z} = 1.
\end{equation}
The rapid rise of density close to the recombination front, \cf\
equation~\refeq{e:struct}, leads to an edge brightened structure in
both line flux and in the far stronger free-free continuum.  The peak
intensity is more than twice that at the centre of the of the region,
for $M|R=2.2$.

In Figure~\ref{f:lineprof}, we show line profiles for an optically
thin recombination line ($I\propto n^2$) at $z=0.1$ and $0.7$.  For
flows with Mach numbers $m\ga2$ at the recombination front, the
predicted line profiles are symmetric but noticeably double-peaked,
particularly in the weaker emission lines close to the centre of the
nebula.  For smaller Mach numbers at the RF, the line profiles will be
singly peaked, although noticeably broadened and variable across an
individual \UCHIIR.  Continuum optical depths may lead to a systematic
blueshift of the emission in lower frequency recombination lines.
Clearly, observations of radio recombination line profiles is a key
test of the model presented here.

\begin{figure}
\centering
\mbox{\epsfxsize=\figwidth\epsfbox{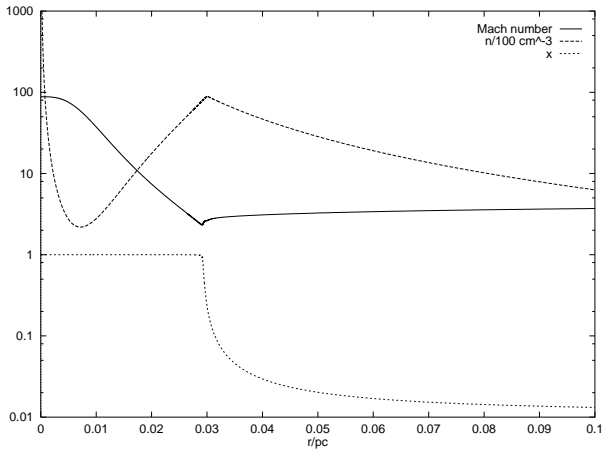}}
\caption{The structure of an isothermal model \UCHIIR, for
parameters given in the text.  The solid curve shows the Mach number
of the flow, the dashed curve the flow density (in units of $100\cm^{-3}$),
and the dotted curve the ionized fraction, $x$.}
\label{f:simplemod}
\end{figure}
In Figure~\ref{f:simplemod}, we show the structure of the \UCHIIR\
calculated for the simple model parameters given above, assuming that
the gas is everywhere isothermal, at $T = 8000\Kelv$ (cf.\ Arthur,
Dyson \& Hartquist 1994).  This is a fairly reasonable assumption
within the ionized nebula.  Outside the recombination front, the gas
would be expected to cool adiabatically; however, since the gas is
supersonic here, the cooling will have little dynamical effect.
Beyond the ionized region, the Mach number plotted should be
interpreted as the flow velocity scaled to the isothermal sound speed.
Note that the neutral hydrogen emerging from the RF has a velocity of
$45\kms$.  Clearly, the observation of neutral material with high
radial velocities close to the projected centre of an \UCHIIR\ would
again constitute a crucial test of such models.

\section{Discussion}

In this paper, we have discussed one of the mechanisms for the
formation of ultracompact \HII\ regions introduced by Dyson~(1994).
Winds around young massive stars mass-loaded by the remaining shreds
of the molecular cloud from which the star initially formed can
naturally explain many of the properties of spherical \UCHIIR.

For parameters characteristic of spherical \UCHIIR, the mass-loaded
stellar wind will remain supersonic at all radii, so long as the
radiating gas has a filling factor $f\la 0.02$.  This filling factor
is close to the preferred range $0.03-0.10$ quoted by Afflerbach
\etal\ (1994) from a non-LTE analysis of recombination line ratios for
the (cometary) \UCHIIR\ G29.96$-$0.02, although limb brightening on a
finer scale than the observational resolution would bias their results
towards small values.

We have shown predicted line profiles and distribution of emission
measure across and \UCHIIR.  The double-peaks of line profiles,
particularly close to the centre of the nebula, are a strong
observational diagnostic, so long as sufficient spatial and spectral
resolution can be obtained.  Observations of sufficient quality are
now beginning to become available for analysis.

In future papers, we will extend the work presented here to treat
centre-bright nebulae (most naturally understood as flows which are
subsonic for a substantial range of radii).  We will also include more
details of the radiation transfer process (such as continuum optical
depths and stimulated emission), of the wind-clump interfaces and of
the distribution and dynamical effects of dust.  Preliminary models
incorporating more realistic heating and cooling rates predict
temperatures appreciably greater than $10^4\Kelv$ in the inner
regions.  This will result in edge brightening even more pronounced
than that shown above (Williams, Dyson \& Redman, in preparation).

Radio recombination lines from helium and carbon, while fainter than
those of hydrogen will give better velocity resolution; the
differences in ionization potential between these atoms means that
comparison of their line profiles will yield important additional
information on the radial structure of the nebula.

\section*{Acknowledgements}

The authors wish thank the anonymous referee, in particular for
suggesting that we extend the discussion of the nature of clumps, and
Robert Massey for discussions on the Trapezium cluster.  This work was
supported by PPARC both through the Rolling Grant to the Astronomy
Group at Manchester (RJRW) and through a Graduate Studentship (MPR).

\label{lastpage}
\end{document}